\documentclass[twocolumn,a4paper]{article-nd2001_2}
\usepackage{times,nd2001_2}
\usepackage[dvips]{graphicx}
\usepackage{jnstcite}
\def\be{\begin{equation}}
\def\ee{\end{equation}}
\def\bea{\begin{eqnarray}}
\def\eea{\end{eqnarray}}

\def\apj #1 #2 #3 {{\it Astrophys. J.}, {\bf #2}, #3, (#1).}
\def\apjl #1 #2 #3 {{\it Astrophys. J. Lett.}, {\bf #2}, L#3, (#1).}
\def\apjs #1 #2 #3 {{\it Astrophys. J. Suppl.}, {\bf #2}, #3, (#1).}
\def\aap  #1 #2 #3 {{\it Aatron. and Astrophys.}, {\bf #2}, #3, (#1).}
\def\mnras #1 #2 #3 {{\it MNRAS}, {\bf #2}, #3, (#1).}
\def\pra #1 #2 #3 {{\it Phys.~Rev.~A.}, {\bf #2}, #3, (#1).}
\def\prb #1 #2 #3 {{\it Phys.~Rev.~B.}, {\bf #2}, #3, (#1).}
\def\prc #1 #2 #3 {{\it Phys.~Rev.~C.}, {\bf #2}, #3, (#1).}
\def\prd #1 #2 #3 {{\it Phys.~Rev.~D.}, {\bf #2}, #3, (#1).}
\def\pre #1 #2 #3 {{\it Phys.~Rev.~E.}, {\bf #2}, #3, (#1).}
\def\prl #1 #2 #3 {{\it Phys.~Rev.~Lett.}, {\bf #2}, #3, (#1).}
\def\plb #1 #2 #3 {{\it Phys.~Lett.~B.}, {\bf #2}, #3, (#1).}
\def\science #1 #2 #3 {{\it  Science.}, {\bf #2}, #3, (#1).}
\def\nature #1 #2 #3 {{\it Nature.}, {\bf #2}, #3, (#1).}
\def\nphysa #1 #2 #3 {{\it Nucl.~Phys.~A.}, {\bf #2}, #3, (#1).}
\def\nphysb #1 #2 #3 {{\it Nucl.~Phys.~B.}, {\bf #2}, #3, (#1).}
\def\nphysbs #1 #2 #3 {{\it Nucl.~Phys.~B.~Suppl.}, {\bf #2}, #3, (#1).}

\def\h#1{\hbox{${}^{#1}$H}}
\def\he#1{\hbox{${}^{#1}$He}}

\def\be#1{\hbox{${}^{#1}$Be}}

\def\omegab{\hbox{${\Omega}_{\rm b}$}}
\def\h502{\hbox{$ h^{2}_{50}$}}
\def\xinue{\hbox{$\xi_{\nu_{e}}$}}
\def\xinum{\hbox{$\xi_{\nu_{\mu}}$}}
\def\xinut{\hbox{$\xi_{\nu_{\tau}}$}}
\def\xinumt{\hbox{$\xi_{\nu_{\mu, \tau}}$}}

\def\ev{\mbox{~eV}}

\def\ga{\mathrel{\mathpalette\fun >}}
\def\fun#1#2{\lower3.6pt\vbox{\baselineskip0pt\lineskip.9pt
  \ialign{$\mathsurround=0pt#1\hfil##\hfil$\crcr#2\crcr\sim\crcr}}}

\begin{document}
\title{\bf Cosmological Nucleosynthesis in the Big-Bang and Supernovae}
\author{ Toshitaka KAJINO$^{1,2,\ast}$\\
\begin{tabular}{c}
\vspace{-3.5mm}
 \\
\fontsize{9}{12}
  $^{1}$\textit{National Astronomical Observatory, Mitaka, Tokyo 181-8588,
Japan} \\
\fontsize{9}{12}
  $^{2}$\textit{University of Tokyo, Graduate School of Science,
  Department of Astronomy, Bunkyo-ku, Tokyo 113-0033, Japan} \\
\end{tabular}
}

\begin{abstract}
Recent observation of the power spectrum of Cosmic Microwave Background (CMB)
Radiation has exhibited that the flat cosmology is most likely.
This suggests too large universal baryon-density
parameter $\Omega_b h^2 \approx 0.022 \sim 0.030$ to accept
a theoretical prediction, $\Omega_b h^2 \le 0.017$, in the homogeneous
Big-Bang model for primordial nucleosynthesis.
Theoretical upper limit arises from the sever constraints
on the primordial $^7$Li abundance.
We propose two cosmological models in order to resolve the descrepancy;
lepton asymmetric Big-Bang nucleosynthesis model,
and baryon inhomogeneous Big-Bang nucleosynthesis model.
In these cosmological models the nuclear processes are
similar to those of the r-process nucleosynthesis in gravitational
collapse supernova explosions.

Massive stars $\geq 10 M_{\odot}$ culminate their evolution by supernova
explosions
which are presumed to be the most viable candidate site for the
r-process nucleosynthesis.  Even in the nucleosynthesis of heavy elements,
initial entropy and density at the surface of proto-neutron stars are so high
that nuclear statistical equilibrium favors production of abundant light
nuclei.
In such explosive circumstances many neutron-rich radioactive nuclei
of light-to-intermediate mass as well as heavy mass nuclei play the
significant roles.

\begin{center} \begin{minipage}{145mm}
\end{minipage}\end{center}

\end{abstract}
\date{}

\maketitle

\keyword[~$^{\ast }$]{Corresponding author, Tel. +81-422-34-3740,
Fax. +81-422-34-3746, E-mail: kajino@nao.ac.jp}

\section{Big-Bang Cosmology}

Recent progress in cosmological deep survey has clarified progressively the
origin and distribution of matter and evolution of Galaxies in the Universe.

The origin of the light elements among them has been a topic of broad interest
for its significance in constraining the dark matter component in the Universe
and also in seeking for the cosmological model which best fits the recent data
of cosmic microwave background (CMB) fluctuations.
This paper is concerned with neutrinos during Big-Bang nucleosynthesis (BBN).
In particular, we consider new insights into the possible role which degenerate
neutrinos may have played in the early Universe~\cite{orito00} .

There is no observational reason to insist that the universal  lepton number
is zero. It is possible, for example,  for the individual lepton
numbers  to be large compared to the baryon number of the
Universe, while the net total lepton number is small $L \sim B$.
It has been proposed recently~\cite{casas99} that models based
upon the Affleck-Dine scenario of baryogenesis might generate naturally lepton
number asymmetry which is seven to ten orders of magnitude
larger than the baryon number asymmetry. Neutrinos with large lepton asymmetry
and masses $\sim 0.07 \ev$ might even explain the
existence of cosmic rays  with energies in excess of
the Greisen-Zatsepin-Kuzmin cutoff \cite{gelmini99}.
It is, therefore, important for both particle physics and cosmology
to carefully scrutinize the limits which cosmology places on the allowed range
of both the lepton and baryon asymmetries.

\subsection{Lepton Asymmetric Big-Bang Model}

Although lepton asymmetric BBN has been studied in many papers~\cite{kang92}
(and references therein), there are several differences in the present work:
For one , we have included finite temperature  corrections to the mass of
the electron
and photon~\cite{fornengo97}.
Another is that we have calculated the neutrino annihilation
rate in the cosmic comoving frame, in which the M{\o}ller velocity
instead of the relative velocity is to be used for the integration of the
collision term in the Boltzmann equations~\cite{gondolo91,enqvist92}.

Neutrinos and anti-neutrinos drop out of thermal equilibrium with the
background thermal plasma  when the weak reaction rate becomes slower than
the universal expansion rate.  If the neutrinos  decouple early, they are
not heated as the particle degrees of freedom change.
Hence, the ratio of the neutrino to photon temperatures, $T_\nu/T_\gamma$,
is reduced.  The biggest drop in temperature for all three neutrino flavors
occurs for $\xi_\nu \sim 10$.    This corresponds to a decoupling
temperature above the cosmic QCD phase transition.
$\xi_\nu$ is the neutrino degeneracy parameter
defined by $\xi_\nu = \mu_\nu/T_\nu$, where $\mu_\nu$ is the chemical potential
and $T_\nu$ is the neutrino temperature.  Finite $\xi_\nu$ leads to
a lepton asymmetric (L $\ne$ 0) Universe.

Non-zero lepton numbers affect nucleosynthesis in two
ways. First, neutrino degeneracy increases the expansion rate.
This increases the  $\he4$ production.
Secondly, the equilibrium n/p ratio is affected by the electron neutrino
chemical potential,
$\rm{n/p} = exp\{-(\Delta \it{M/T}_{n \leftrightarrow p}) - \xinue\}$,
where $\Delta M$ is the neutron-proton mass difference and
$T_{n \leftrightarrow p}$ is the freeze-out temperature for the relevant
weak reactions.
This effect either increases or decreases $\he4$ production,
depending upon the sign of $\xinue$.

A third effect emphasized in this paper is that
$T_\nu/T_\gamma$ can be reduced if the neutrinos decouple
early.  This lower temperature reduces the energy density of
neutrinos during BBN, and slows the expansion of the
Universe. This decreases $\he4$ production.

{\bf Figure 1} highlights the main result of this study,
where we take $\xinum = \xinut$.  For low $\omegab \h502 $ models,
only the usual low values for $\xinue $ and $\xinumt$ are allowed.
Between $\omegab \h502 \approx$ 0.188 and 0.3, however, more
than one allowed region emerges.
For $\omegab \h502 \ga 0.4$ only the large degeneracy solution
is allowed.  Neutrino degeneracy can even allow baryonic densities
up to $\omegab \h502 = 1$.

\begin{figure}[h]
 \begin{center}
 \includegraphics[width=8.6cm,clip]{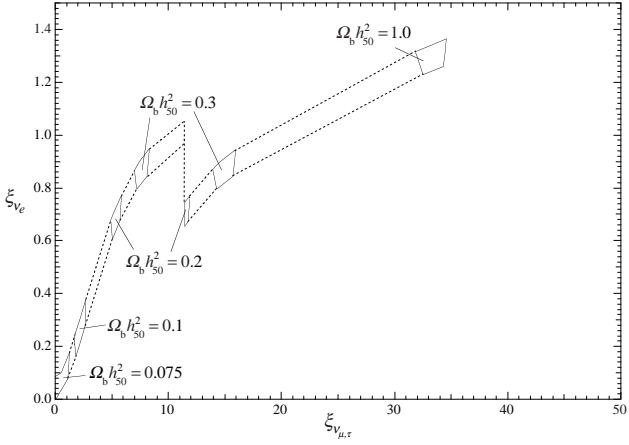}
   \caption{Allowed values of $\xinue$ and $\xinumt$ for which the
constraints from light element abundances are satisfied for
values of $\omegab \h502 =$ 0.075, 0.1, 0.2, 0.3 and 1.0 as indicated.
Note that $\omegab \h502 = 4 \times \omegab h^2$, and $h = h_{100}$.}
   \label{fig:fig8}
 \end{center}
\end{figure}

\subsection{Cosmic Microwave Background}

Several recent works~\cite{kinney,lesgourgues,hannestad} have shown that
neutrino degeneracy can dramatically alter the power spectrum of the CMB.
However, only small degeneracy parameters with the standard relic neutrino
temperatures have been utilized.  Here, we have calculated the CMB power
spectrum
to investigate effects of a diminished relic neutrino temperature.

The solid line on {\bf Fig. 2} shows a  $\Omega_\Lambda = 0.4$
model for which $n = 0.78$, where n is the power index of primordial
fluctuations.
This fit is marginally consistent with the data at a level of $5.2 \sigma$.
The dotted line shows the matter dominated $\Omega_\Lambda = 0$ best fit model
with $n = 0.83$ which is consistent with the data at the level of $3 \sigma$.
The main differences in the fits between the large degeneracy models
and our adopted benchmark model are that the first peak is shifted to slightly
higher $l$ value and the second peak is suppressed.
One can clearly see that the suppression of the second acoustic peak is
consistent
with our derived neutrino-degenerate models.
In particular, the MAXIMA-1 results are in very good agreement with the
predictions of our neutrino-degenerate cosmological models.
It is clear that these new data sets substantially improve the
goodness of fit for the neutrino-degenerate models~\cite{lesgourgues}.
Moreover, both data sets seem to require an increase in
the baryonic contribution to the closure density
as allowed in our neutrino-degenerate models~\cite{orito00,mathews01}.

\begin{figure}[h]
 \begin{center}
 \includegraphics[width=8.6cm,clip]{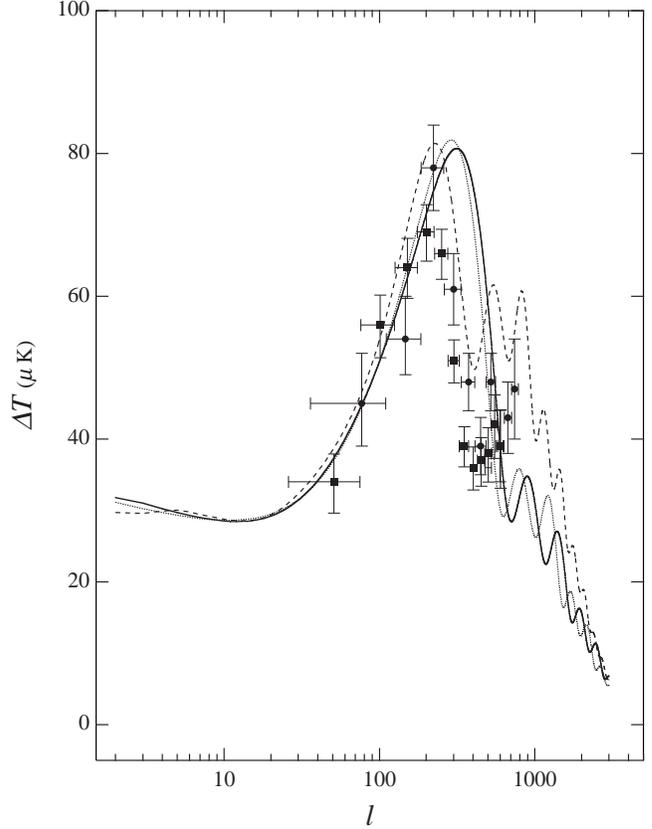}
   \caption{CMB power spectrum  from
BOOMERANG~\protect\cite{boomerang} (squares) and
MAXIMA-1~\protect\cite{hanany}
(circles) binned data compared with calculated $\Omega = 1$ models.}
   \label{fig:fig15}
 \end{center}
\end{figure}

\subsection{Baryon Inhomogeneous Big-Bang Model}

The biggest advantage of the baryon inhomogeneous Big-Bang nucleosynthesis
model~\cite{applegate87,alcock87,kajino90} is to allow larger $\omegab h^2
\le 0.05$,
which well covers the constraint from recent CMB data
$\Omega_b h^2 \approx 0.022 \sim 0.030$, still satisfying the light element
abundance constraints.

Let us consider what kind of observational signature is expected in this model.
Nuclear reaction flow stops at the A = 7 nuclear systems in the homogeneous
Big-Bang nucleosynthesis model because of the instability of $^8$Be.
In the baryon inhomogeneous model, however,
the nucleosynthesis occurs in an environment of proton-neutron segregated
inhomogeneous distribution.
Therefore, the radioactive nuclear reactions play the significant roles
in the production of intermediate-to-heavy mass elements
via unstable nuclei $^8$Li(838 ms), $^9$Li(178.3 ms), $^7$Be(53.29 d),
$^{10}$Be($1.51 \times 10^6 y$), $^8$B(770 ms), $^{12}$B(20.20 ms),
$^{11}$C(20.385 m), $^{14}$C(5730 y), $^{15}$C(2.449 s),
$^{13}$N(9.965 m), $^{16}$N(7.13 s), $^{14}$O(70.606 s), etc.
\begin{eqnarray}
&^4\rm{He}(^3\rm{H},\gamma)^7\rm{Li}(n,\gamma)^8\rm{Li}(\alpha,n)^{11}\rm{B}
(n,\gamma)
^{12}\rm{B}(\beta~\nu) \\ \nonumber
&^{12}\rm{C}(n,\gamma)^{13}\rm{C}(n,\gamma)^{14}\rm{C} ... ,
\end{eqnarray}
\begin{eqnarray}
&^7\rm{Li}(n,\gamma)^8\rm{Li}(n,\gamma)^9\rm{Li}(\beta~\nu)^9\rm{Be}(n,
\gamma)^{10}\rm{Be}(n,\gamma) \\ \nonumber
&^{11}\rm{Be}(\beta~\nu)^{11}\rm{B} ... ,
\end{eqnarray}
\begin{eqnarray}
&^7\rm{Li}(^3\rm{H},n)^9\rm{Be}(^3\rm{H},n)^{11}\rm{B}.
\end{eqnarray}
The two reaction chains (1) and (2) play the key roles in the production of
heavy neutron-rich isotopes~\cite{applegate87,kajino90} in the neutron-rich
zones.
Since $^7$Li is the heaviest element to be created in the homogeneous Big-Bang
nucleosynthesis model, all nuclear reactions for the production of heavier
elements
have ever been ignored in the previous calculations.  We however found
that the reaction chain (3) is extremely important for the production of
$^9$Be in the baryon inhomogeneous Big-Bang nucleosynthesis
models~\cite{kajino89,kajino90be,orito97}.
With these reactions being included in the network of the baryon inhomogeneous
Big-Bang models, the $^9$Be abundance increases by three orders of magnitude to
N(Be)/N(H) $\approx 10^{-14}$ which approaches the current observational level.


\section{Supernova Explosion}

Stars with various masses provide a variety of production sites
for intermediate-to-heavy mass elements.  Very massive stars
$\geq 10 M_{\odot}$ culminate their evolution by supernova (SN) explosions
which are presumed to be most viable candidate for the still unknown
astrophysical site of r-process nucleosynthesis.
We discuss in this section the neutrino-driven winds from Type II SN
explosion of very massive stars.
Although there is still a room for the prompt explosion~\cite{sumiyoshi01}
to account for one part of the r-process nucleosynthesis,
we concentrate on the gravitaional core-collpase Type II SNe here.

Even in the nucleosynthesis of heavy elements, initial entropy and density
at the surface of proto-neutron stars are so high that nuclear statistical
equilibrium (NSE) favors production of abundant light nuclei.
In such explosive circumstances of so called hot-bubble scenario,
not only heavy neutron rich nuclei but light unstable
nuclei play a significant role.

The study of the origin of r-process elements is also critical in cosmology.
It is a potentially serious problem that the cosmic age of the expanding
Universe
derived from cosmological parameters may be shorter than the age of
the oldest globular clusters.  Since both age estimates are subject to the
uncertain
cosmological distance scale, an independent method has long been needed.
Thorium, which is a typical r-process element and has half-life of 14 Gyr,
has recently been detected along with other elements in very metal-deficient
stars.
If we model the r-process nucleosynthesis in these first-generation
stars, thorium can be used as a cosmochronometer
completely independent of the uncertain cosmological distance scale.

\subsection{Neutrino-Driven Winds in Type-II Supernovae}

Recent measurements using high-dispersion spectrographs with large
Telescopes or the Hubble Space Telescope have made it possible
to detect minute amounts of heavy elements in faint metal-deficient
([Fe/H] $\le$ -2) stars~\cite{sneden96}. The discovery of r-process elements
in these
stars has shown that the relative abundance pattern for the mass region
120 $\le$ A is surprisingly similar to the solar system r-process abundance
independent of the metallicity of the star.
Here metallicity is defined by
[Fe/H] = log[N(Fe)/N(H)] - log[N(Fe)/N(H)]$_{\odot}$. It obeys the
approximate relation t/10$^{10}$yr $\sim$ 10$^{[Fe/H]}$.
The observed similarity strongly suggests that the r-process occurs
in a single environment which is independent of progenitor
metallicity.  Massive stars with 10$M_{\odot} \le M$ have a short life
$\sim 10^7$ yr and eventually end up as violent supernova explosions,
ejecting material into the intersteller medium early on quickly from the
history of the Galaxy.
However, the iron shell in SNe is excluded from being the
r-process site because of the observed metallicity independence.

Hot neutron stars just born in the gravitational core collapse
SNeII release most of their energy as neutrinos during the
Kelvin-Helmholtz cooling phase.  An intense flux of neutrinos heat the
material near
the neutron star surface and drive matter outflow (neutrino-driven winds).
The entropy in these winds is so high that the NSE favors a plasma
which consists of mainly free nucleons and alpha particles rather than
composite nuclei like iron.  The equilibrium lepton fraction,
$Y_e$, is determined by a delicate balance between
$\nu_e + n \rightarrow p + e^-$ and $\bar{\nu}_e + p \rightarrow n + e^+$,
which overcomes the difference of chemical potential between $n$ and $p$,
to reach $Y_e \sim$ 0.45.  R-process nucleosynthesis occurs because
there are plenty of free neutrons at high temperature.
This is possible only if seed elements are produced in the correct
neutron-to-seed ratio before and during the r-process.

Although Woosley et al.~\cite{woosley94} demonstrated a profound
possibility that the r-process could occur in these winds, several
difficulties
were subsequently identified.  First, independent non relativistic
numerical supernova models~\cite{witti94} have difficulty producing
the required entropy in the bubble S/k $\sim$ 400.
Relativistic effects may not be enough to increase
the entropy dramatically~\cite{qian96,cardall97,otsuki00}.
Second, even should the entropy be high enough, the effects of neutrino
absorption $\nu_e + n \rightarrow p + e^-$ and
$\nu_e + A(Z,N) \rightarrow A(Z+1,N-1) + e^-$
may decrease the neutron fraction during the nucleosynthesis process.
As a result, a deficiency of free neutrons could prohibit the
r-process~\cite{meyer95}.

In order to resolve these difficulties, we have
studied~\cite{otsuki00,wanajo01}
neutrino-driven winds in a Schwarzschild geometry under the
reasonable assumption of spherical steady-state flow.
The parameters in the wind models are the mass of neutron star, $M$,
and the neutrino luminosity, $L_{\nu}$.
The entropy per baryon, S/k, in the asymptotic regime and the expansion
dynamic time scale, $\tau_{dyn}$, which is defined
as the duration time of the $\alpha$-process when the temprature drops from
T $\approx$ 0.5 MeV to 0.5/e MeV,
are calculated from the solution of hydrodynamic equations.
Then, we carried out r-process nucleosynthesis calculations in our wind model.
We found~\cite{otsuki00} that the general relativistic effects make
$\tau_{dyn}$
much shorter, although the entropy increases by about 40 \% from the
Newtonian value of S/k $\sim$ 90.
By simulating many supernova explosions, we have found some interesting
conditions
which lead to successful r-process nucleosynthesis,
as to be discussed in the following sections.

\subsection{R-process Nucleosynthesis}

Previous r-process calculations~\cite{woosley94,meyer92} had complexity that
the
seed abundance distribution was first calculated by using
smaller network for light-to-intermediate mass elements,
and then the result was connected further to another r-process network
in a different set of the computing run.
For this reasaon it was less transparent to interpret the whole
nucleosynthesis process.
This inconvenience happened because it was numerically too heavy to run
both $\alpha$-process and r-process in a single network code for too huge
number of reaction couplings among $\sim 3000$ isotopes.
Our nucleosynthesis calculation~\cite{otsuki00,wanajo01}
is completely free from this complexity
because we exploited fully implicit single network code which
is applied to a sequence of the whole processes of NSE - $\alpha$-process -
r-process.

\begin{center}
\begin{figure}[t]
\includegraphics[width=60mm,angle=-90]{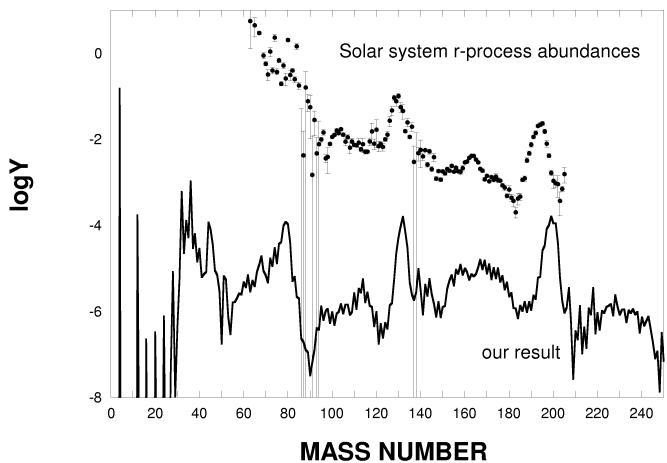}
\caption{
R-process abundance~\protect\cite{otsuki00} (solid line) as a
function of atomic mass number
$A$ compared with the solar system r-process abundance (filled
circles)
from K\"appeler, Beer, \& Wisshak~\protect\cite{kappeler89}.
The neutrino-driven wind model used is for
$L_{\nu} = 10^{52}$ ergs/s and $M = 2 M_{\odot}$.
The solar system r-process abundance is shown in arbitrary unit.
\label{abundance1}}
\end{figure}
\end{center}

Let us remind the readers that there were at least three difficulties
in the previous theoretical studies of the r-process.
The first difficulty among them is that an ideal, high entropy
in the bubble S/k $\sim$ 400~\cite{woosley94} is hard to be achieved in
the other simulations~\cite{witti94,qian96,cardall97,otsuki00}.

The key to resolve this difficulty is found with the short dynamic time scale
$\tau_{dyn}\sim$ 10 ms in our models~\cite{otsuki00,wanajo01}
of the neutrino-driven winds.  As the initial nuclear
composition of the relativistic plasma consists of neutrons and protons,
the $\alpha$-burning begins when the plasma temperature cools below
T $\sim$ 0.5 MeV.  The $^4$He$(\alpha \alpha,\gamma)^{12}$C reaction
is too slow at this temperature, and alternative nuclear reaction path
$^4$He$(\alpha n,\gamma)^9$Be$(\alpha,n)^{12}$C triggers explosive
$\alpha$-burning to produce seed elements with A $\sim$ 100~\cite{woosley92}.
Therefore, the time scale for nuclear reactions is regulated by
the $^4$He$(\alpha n,\gamma)^9$Be. It is given by
$\tau_N \equiv \left(\rho_b^2 Y_{\alpha}^2 Y_n \lambda(\alpha \alpha n
\rightarrow ^9{\rm Be}) \right)^{-1}$.
If the neutrino-driven winds fulfill the condition
$\tau_{dyn} < \tau_N$, then fewer seed nuclei are produced
during the $\alpha$-process with plenty of free neutrons left over when
the r-process begins at T $\sim$ 0.2 MeV.  The high neutron-to-seed
ratio, $n/s \sim 100$, leads to appreciable production of r-process
elements, even for low entropy S/k $\sim$ 130, producing both the 2nd
$(A \sim 130)$ and 3rd $(A \sim 195)$ abundance peaks and the hill of
rare-earth elements $(A \sim 165)$ ({\bf Fig. \ref{abundance1}}).

\begin{figure}[h]
 \begin{center}
 \includegraphics[width=8.6cm,clip]{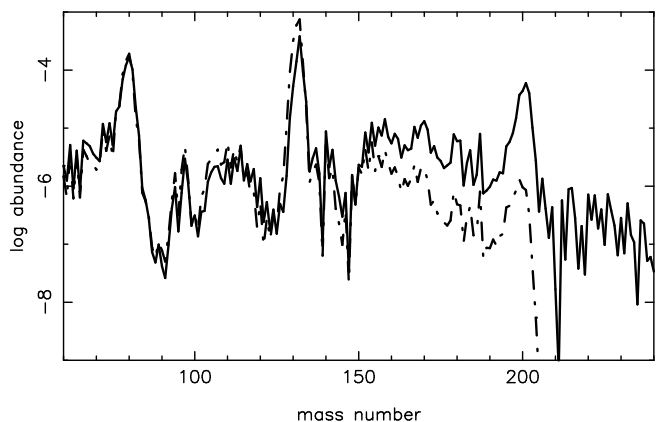}
   \caption{The same as those in Figure \protect\ref{abundance1},
but for the neutrino-driven wind model of
$L_{\nu} = 5 \times 10^{52}$ ergs/s.
Solid line respresents the result by using the Woosley \& Hoffman
rate~\protect\cite{woosley92} of the  $^4$He$(\alpha n,\gamma)^9$Be
reaction,
and long-dashed line for the rate multiplied by factor 2,
as suggested by the recent experiment of
Utsunomiya et al.~\protect\cite{utsunomiya01}.}
   \label{fig:abundance2}
 \end{center}
\end{figure}
%

The three body nuclear reaction cross section for $^4$He$(\alpha
n,\gamma)^9$Be
is one of the poorly determined nuclear data which may
alter the r-process nucleosynthesis yields.
The inverse process has recently been studied experimentally
by Utsunomiya et al.~\cite{utsunomiya01},
and photodisintegration cross section of $^9$Be
has been measured with better precision than those of the previous experiments.
Applying the principle of the detailed balance to this process, one can
estimate the
cross section for $^4$He$(\alpha n,\gamma)^9$Be.
They found that the thermonuclear reaction rate is almost twice as big as
that of
Woosley and Hoffman~\cite{woosley92} but in resonable agreement with recent
compilation of Angulo et al.~\cite{angulo99}.
However, there still remain several questions on the consistency between
their result and electron-scattering experiments, on the contribution
from the narrow resonance $J^{\pi} = 5/2^{-}$ (2.429 MeV), etc.
It is also a theoretical challenge to understand the reaction
mechanism and the resonance structure because two different channels,
$^8$Be + n and $^5$He + $\alpha$, contribute to this process.

Therefore, we show two calculated results in {\bf Fig. 4}:
The solid line displays the result obtained by using the Woosley and Hoffman
cross section~\cite{woosley92}, assuming a $^8$Be + n structure
for $^9$Be.  We also calculated the r-process by multiplying
this cross section by factor of 2 (long-dashed line).  This makes a
drastic change in the r-process yields in the 3rd $(A \sim 195)$ abundance
peak.
More theoretical and experimental studies of the
$^4$He$(\alpha n, \gamma)^9$Be reaction are highly desired.

\subsection{Neutrino-Nucleus Interactions}

Neutrino interactions with nucleons and nuclei take the key to resolve
the second difficulty which was pointed out in sect. 1.  The difficulty is that
the effects of neutrino absorptions $\nu_e + n \rightarrow p + e^-$ and
$\nu_e + A(Z,N) \rightarrow A(Z+1,N-1) + e^-$
during the $\alpha$-process may induce the deficiency of free neutrons
and break down the r-process conditions~\cite{meyer95}.
These two types of neutrino interactions control most sensitively
the electron fraction and the neutron fraction, as well, in a neutron-rich
environment.
In order to resolve this difficulty, we have updated the electron-type
neutrino capture
rates for all nuclei and electron-type anti-neutrino capture
rate for free proton~\cite{qian97,meyer98}.

The new r-process calculation proves to be almost invariant.
One can understand this robustness of the succesful r-process in the
following way:
The specific collision time for neutrino-nucleus interactions is given by
\begin{equation}
\tau_{\nu} \approx 201 \times L_{\nu ,51}^{-1} \times
\left(\frac{\epsilon_{\nu}}{\rm MeV}\right)
\left(\frac{r}{\rm 100km}\right)^2
\left(\frac{\langle\sigma_{\nu}\rangle}{\rm 10^{-41}cm^2}
\right)^{-1} ms,
\label{eqn:tau}
\end{equation}
where $L_{i,51}$ is the individual neutrino or
antineutrino luminosity in units of $10^{51}$ ergs/s,
$\epsilon_i=<E_i^2>/<E_i>$ in MeV $(i=\nu_e,~~\bar{\nu}_e,~~etc. )$,
and $\langle \sigma _{\nu} \rangle$ is the averaged cross
section over neutrino energy spectrum.
At the $\alpha$-burning site of r $\approx$ 100 km
for $L_{\nu ,51} \approx 10$, $\epsilon_{\nu_e}=12~\rm{MeV}$,
and $\langle \sigma _{\nu} \rangle \approx 10^{-41} cm^2$,
$\tau_{\nu_e}$(r=100 km) turns out to be $\approx 240$ ms.
This collision time is larger than the expansion dynamic time scale;
$\tau_{dyn} \approx$ 10 ms $\ll$ $\tau_{\nu_e}$(r=100 km) $\approx
240$ ms.  Because there is not enough time for $\nu_e$'s
to interact with n's in such rapidly expanding neutrino-driven wind,
the neutron fraction is insensitive to the neutrino absorptions.

One might wonder if our dynamic time scale $\sim 10$ ms
is too short for the wind to be heated by neutrinos.
Careful comparison between proper expansion time and specific
collision time for the neutrino heating is needed in order to
answer this question.  Otsuki et al.~\cite{otsuki00}
have found that the supernova neutrinos transfer their
kinetic energy to the wind most effectively just above the neutron star
surface at $10 {\rm km} \leq r < 20 {\rm km}$.
Therefore, one should refer the duration
time for the wind to reach the $\alpha$-burning site, $\tau_{\rm heat}$,
rather than $\tau_{\rm dyn}$.  One can estimate this heating time scale
\begin{equation}
\tau_{\rm heat} =
\int^{r_f}_{r_i}\frac{dr}{u},
\label{eqn:timeheat}
\end{equation}
where u is the fluid velocity of the wind.  By setting the radius of
neutron star surface $r_i=10$ km and $r_f=100 $ km,
we get $\tau_{\rm heat} \approx$ 30 ms.
The collision time $\tau_{\nu}$ is
given by Eq.~(\ref{eqn:tau}) by setting $L_{\nu ,51} \approx 10$,
$\epsilon _{\nu}=(\epsilon _{\nu_ e} + \epsilon _{\bar{\nu}_e} )/2
= (12 + 22)/2 = 17$ MeV,
r $\approx$10 km, and $\langle \sigma _{\nu} \rangle \approx 10^{-41} cm^2$.
Let us compare $\tau_{\rm heat}$ and $\tau_{\nu}$ to one another:
%
$\tau_{\nu} \approx 3.4 {\rm ms} ~\ll~ \tau_{\rm heat} \approx 30 {\rm
ms}.$
%
We can thus conclude that there is enough time for the expanding wind to be
heated by neutrinos even with short dynamic time scale
for the $\alpha$-process, $\tau_{\rm dyn} \sim 10$ ms.

\subsection{Roles of Light Neutron-Rich Nuclei}

The r-process is thought to proceed after the pile up of seed nuclei
produced in the $\alpha$-process at higher temperatures
$T_9 \approx 5 \sim 2.5$.
Since charged-particle reactions, which reassemble nucleons into
$\alpha$-particles
and $\alpha$-particles into heavier nuclei (i.e. $\alpha$-process),
are faster than the neutron-capture flow which is regulated by beta-decays,
the light-mass neutron-rich nuclei were presumed to be unimportant.

However, Terasawa et al.~\cite{terasawa01} have recently found that even
light neutron-rich
nuclei progressively play the significant roles in the production of seed
nuclei.
Nuclear reaction network used in the previous studies~\cite{woosley94,meyer92}
includes only limited number of light unstable nuclei, $^{3}$H, $^{7}$Be,
$^{8,9}$B,
$^{11,14}$C, $^{13}$N, $^{15}$O, $^{18,20}$F, $^{23,24}$Ne, and so on.
We therefore need to extend the network code so that it covers
all radioactive nuclei to the neutron-drip line.
We take the rates of charged particle reactions from those used in the
Big-Bang nucleosynthesis calculations~\cite{kajino90,kajino90be,orito97}
and the NACRE compilation~\cite{angulo99}.

Let us briefly discuss preliminary result of the r-process calculation,
using the extended reaction network~\cite{terasawa01}.
At early epoch of the wind expansion, t $\le$ a few dozens ms,
both temperature and density are so high that the charged particles
interact with one aother to proceed nucleosynthesis around the
$\beta$-stability line
in the light-mass region A $\le$ 20.
There are plenty of protons and $\alpha$-particles as well as neutrons
at this epoch, and the main reaction flow is triggered by
$^4\rm{He}(\alpha n,\gamma)^9\rm{Be}$~\cite{woosley94,woosley92}:
\begin{eqnarray}
&^4\rm{He}(\alpha
n,\gamma)^9\rm{Be}(\alpha,n)^{12}\rm{C}(n,\gamma)^{13}\rm{C}(\alpha,n) \\
\nonumber
&^{16}\rm{O}(n,\gamma)^{17}\rm{O}(\alpha,n)^{20}\rm{Ne}~~or~~^{16}\rm{O}(\alpha,\gamma)^{20}\rm{Ne} ...
\end{eqnarray}
However, at relatively later epoch even after the $\alpha$-rich freeze out,
a new reaction path~\cite{terasawa01}
\begin{eqnarray}
&^3\rm{H}(\alpha,\gamma)^7\rm{Li}(n,\gamma)^8\rm{Li}(\alpha,n)^{11}\rm{B}(n,
\gamma)^{12}\rm{B}(n,\gamma) \\ \nonumber
&^{13}\rm{B}(n,\gamma)^{14}\rm{B}(n,\gamma)^{15}\rm{B}(e^- \nu)^{15}\rm{C} ...
\end{eqnarray}
also takes some appreciable flux of baryon number to continuously supply
the seed nuclei.  The classical r-process like flow, (n,$\gamma$)
followed by beta decay, has already started from light nuclei.
This is a very different result from the previous picture that
the r-process starts from only intermediate-mass seed nuclei $A \approx 100$.

Since we do not have much information of $(2n,\gamma)$ reactions,
we did not include $^{6}$He, $^{8}$He, $^{11}$Li, $^{14}$Be, $^{17,19}$Be,
$^{22}$C, etc.
The yields of even the most neutron-rich isotopes were found to be
abundant in this calculation~\cite{terasawa01}, and
we plan to study the possible role of the (2n,$\gamma$) reactions.
There are several branching points between (n,$\gamma$) and
$(\alpha$,n) reactions.  They are at $^{18}\rm{C}$,
$^{24}\rm{O}$, $^{36}\rm{Mg}$, etc.
Exerimetal studies to measure these reaction cross sections are highly
desirable.

\section{Quest for Nuclear Data and Astrophysics Data}

Our result in Fig. 3 reproduces fairely well the observed
second abundance peak (A $\approx$ 130), the hill of rare-earth elements (A
$\approx$ 165),
and the third peak (A $\approx$ 195).
However, there are several defects, too.
The first defect is a shift of the third abundance peak around A $\approx$ 195
by a couple of mass units.  This is a common feature found in the previous
r-process calculations~\cite{woosley94,otsuki00,meyer92}, too.
These elements are the beta-decay products of extremely neutron-rich
unstable nuclei on the neutron magic N = 126.
Peak position depends on the timing of freezeout of the r-process.
Therefore, a particular combination of environmental evolution of
neutron-number density, $N_n$,
and temperature, $T_9$, as well as the expansion dynamic time scale,
$\tau_{dyn}$, might match the freeze out so that it results in the
right position of abundance peak~\cite{terasawa01}.

The second defect is the deficiency of abundance
right above or below the peak elements,
i.e. at A $\approx$ 90, 120, 150, 190, and 210.
These deficiencies seem related to yet unseen effects of deformation
or strucure change of unstable nuclei surrounding the neutron magic
numbers N = 50, 82, and 126.
Further extensive theoretical studies and observational challenge
to determine the masses, lives, and beta Q-values of these nuclei are highly
desired.

The third defect is the underproduction of actinoid elements,
Th-U-Pu (A = 230 $\sim$ 240), by more than one order of magnitude.
The observed high abundance level of these nuclei
might suggest an existence of a new magic number
around N = $150 \sim 160$:
Xenon $^{129}$Xe$_{75}$ and platinum $^{195}$Pt$_{117}$ are the typical
r-process elements on the second and third abundance peaks,
which are decay products from extremely neutron-rich unstable
nuclei with neutron magic numbers N = 82 and 126, respectively.
From these observations we estimate that the waiting point nucleus
is located by shifting $\Delta$N $\approx$ 7 or 9 units from the peak element.
Applying the same shift $\Delta$N $\approx$ 7 or 9 to $^{232}$Th$_{142}$ and
$^{238}$U$_{146}$, we could assume a new magic number
around N = $150 \sim 160$ which may lead to
the fourth abundance peak at A = 230 $\sim$ 240.
Actually, in the very light nuclear systems, a new magic number N = 16
was found~\cite{ozawa00} in careful experimental studies of the neutron
separation energies and interaction cross sections of extremely neutron-rich
nuclei.
Since these possibilities were not taken into account
in the present and previous calculations,
the deficiency of actinoids might be improved by modernizing nuclear mass
formula
including such effects.
Another possibility is to make actinoid elements in neutron star mergers
or the mergers of the neutron star and black hole binaries which have
extremely small lepton fraction, $Y_e \le 0.2$~\cite{freiburghaus99}.
However, these processes do not virtually produce any
intermediate-mass nuclei including iron,
which contradicts with the fact that the observed iron abundance
is proportional to the r-process and actinoid elements
over the entire history of Galactic evolution $\sim 10^{10}$ yr.

Since we discuss only material ejected from the proto-neutron star
behind the shock, it does not make any serious problem
to see the underproduction in mass region A $\le$ 90.
Most of these intermediat-mass nuclei are ejected from the
exploded outer shells in supernovae.

\section*{{\fontsize{10}{12}\bf Acknowledgment}}
This paper has been supported in part by the
Grant-in-Aid for Science Researches 10044103, 10640236, 12047233
and 13640313 of the Ministry of Education, Science, Sports, and Culture of
Japan.


\setlength{\unitlength}{1mm}
\begin{picture}(81,10)
 \thicklines
 \put(-47.5,8){\line(1,0){81}}
\end{picture}

\end{document}